\documentclass[conference]{IEEEtran}
\IEEEoverridecommandlockouts
\usepackage{caption}
\usepackage{subfigure}
\usepackage{graphics} 
\usepackage{epsfig} 
\usepackage{amsfonts,amssymb}
\usepackage{multirow}
\usepackage{amsmath}
\graphicspath{{figure/}}
\usepackage{subfigure}
\usepackage{graphics}
\usepackage{multirow}
\usepackage{xcolor}
\usepackage{color}

\usepackage{algorithm}
\usepackage{algpseudocode}

\begin{document}

\title{\LARGE \bf  Generalizing Deep Learning-Based CSI Feedback in Massive MIMO via ID-Photo-Inspired Preprocessing\\

\thanks{Z. Liu, Y. Ma, and R. Tafazolli are with the 5GIC and 6GIC, Institute for Communication Systems, University of Surrey, United Kingdom (e-mail: \{zhenyu.liu, y.ma, r.tafazolli\}@surrey.ac.uk). This work has been filed for patent under number 92066794PCT01.}
}

%
%
%
\author{Zhenyu Liu, Yi Ma, and Rahim Tafazolli}


\maketitle

\begin{abstract}

Deep learning (DL)-based channel state information (CSI) feedback has shown great potential in improving spectrum efficiency in massive MIMO systems. However, DL models optimized for specific environments often experience performance degradation in others due to model mismatch.
To overcome this barrier in the practical deployment, we propose UniversalNet, an ID-photo-inspired universal CSI feedback framework that enhances model generalizability by standardizing the input format across diverse data distributions. Specifically, UniversalNet employs a standardized input format to mitigate the influence of environmental variability, coupled with a lightweight sparsity-aligning operation in the transformed sparse domain and marginal control bits for original format recovery. This enables seamless integration with existing CSI feedback models, requiring minimal modifications in preprocessing and postprocessing without updating neural network weights.
Furthermore, we propose an efficient eigenvector joint optimization method to enhance the sparsity of the precoding matrix by projecting the channel correlation into the eigenspace, thus improving the implicit CSI compression efficiency. Test results demonstrate that UniversalNet effectively improves generalization performance and ensures precise CSI feedback, even in scenarios with limited training diversity and previously unseen CSI environments.

\end{abstract}


%
\IEEEpeerreviewmaketitle

\section{Introduction}

Modern wireless communication systems have advanced significantly by utilizing multiple-input multiple-output (MIMO) transceivers, with massive MIMO being essential for 5G and beyond due to its improvements in spectrum and energy efficiency. However, the large antenna arrays and bandwidth in massive MIMO systems increase the volume of feedback data, driving research into efficient downlink channel state information (CSI) feedback methods in frequency division duplex (FDD) systems. Deep learning (DL)-based CSI feedback methods outperform traditional techniques in recovery accuracy and precoding efficiency \cite{Overview}, leveraging channel characteristics such as spatial and spectral correlation \cite{cui2022, mourya2023}, bidirectional reciprocity \cite{ref:canet}, and temporal correlation \cite{liu2022} for efficient full CSI recovery. Additionally, DL-based implicit CSI feedback methods \cite{eigennet_liu, eigennet_xiao, mixeddata2024_2}, aligned with 3GPP-defined codebooks using eigenvectors for partial CSI feedback, have shown better performance than the enhanced Type II (eTypeII) codebook \cite{3gpp2022}. Importantly, the inclusion of a new study item (SI) on ``Artificial Intelligence / Machine Learning for NR Interface'' in 3GPP Release 18 \cite{3gpp2023csi} underscores the potential of DL-based CSI feedback as a leading use case.

One of the key challenges in DL-based CSI feedback is that models tailored to specific radio-frequency (RF) environments often degrade in performance when deployed in different settings due to model mismatch, highlighting the need for better generalization. Two primary strategies have been explored to tackle this issue: model switching using a multiple-encoders-to-multiple-decoders framework, and training enhancement with mixed datasets.

On the one hand, the multiple-encoders-to-multiple-decoders framework \cite{ref:multitask_csi} trains distinct networks for different scenarios, with user equipment (UE) dynamically switching between them based on the detected channel environment. However, this strategy requires substantial memory for storing numerous encoder networks or bandwidth for downloading new encoder parameters, each potentially consisting of millions of parameters \cite{cui2022, mourya2023}. A more optimized solution uses a shared encoder with task-specific decoders to minimize UE storage and model update costs \cite{chen2023oneside}. Nonetheless, the use of a shared encoder compromises CSI recovery accuracy due to the omission of environment-specific features.

On the other hand, the second strategy focuses on improving model generalization by training on datasets from mixed environments \cite{ mixeddata2024, mixeddata2024_2}. Although such models often outperform those trained in single environments, they still suffer performance degradation compared to environment-specific models. Furthermore, while complex models developed from mixed datasets—such as those used in AI competitions with encoders reaching hundreds of megabytes \cite{sis_xiao}—can deliver high performance, they are impractical for energy-constrained and memory-constrained mobile devices.

To address the generalization challenge in the practical deployment of DL-based CSI feedback, we propose an ID-photo-inspired universal CSI feedback solution named UniversalNet. Unlike existing works, UniversalNet draws inspiration from the domain of digital identification, where ID photos undergo predefined preprocessing steps to ensure their universal applicability and unique identification capabilities globally. To the best of our knowledge, this is the first work that utilizes preprocessing and postprocessing without any updates to the neural networks in unseen environments to enhance the generalization capability of DL-based CSI feedback.

Our contributions are summarized as follows:
\begin{itemize}
\item To enhance the generalization capacity of DL-based CSI feedback, instead of optimizing the design and training of neural networks, we propose the creation of a standardized input format to ensure a consistent basis for variable input data distributions. To conform the input CSI to this standardized format with low complexity, we introduce a benchmark for the standard input and a fast sparsity-aligning operation designed to effectively bridge the environmental variability gap. Our proposed framework can seamlessly integrate with existing CSI feedback models by inserting a preprocessing and postprocessing module.

\item We observe that the randomness of the eigenvectors highly disrupts the consistency of physical environment information in implicit CSI feedback, degrading the sparsity of the precoding matrix in the transformed domain and complicating its compression. Based on this insight, we design an efficient eigenvector joint optimization method by projecting the spatial and spectral correlation of the massive MIMO channel into the eigenspace to enhance the sparsity and compression efficiency of precoding matrices further.

\item We evaluate the performance of our method using a ray-tracing-based dataset \cite{WirelessAI2022}, which consists of 100 maps from major cities around the world. By selecting samples from the last 30 maps as the testing set and samples from the first 1--70 maps as the training set, we demonstrate the superiority of our proposed method. Notably, the model trained with data from only one map can achieve a Squared Generalized Cosine Similarity (SGCS) close to 0.9 with 45 feedback bits in unseen 30 maps. Additionally, we present a visualization of our preprocessing technique, illustrating its operational mechanics.
\end{itemize}

\section{System Description}

\subsection{System Model}
This work considers a massive MIMO system at the base station (BS) equipped with $N_t = N_h \times N_v$ transmit antennas, where $N_h$ and $N_v$ denote the horizontal and vertical antenna elements, respectively. The UE is equipped with $N_r$ receive antennas. We define $K$ as the number of subbands, each consisting of $N_{gran}$ subcarriers. The downlink channel matrix, $\mathbf{H}'$, is represented as:
\begin{equation}
\mathbf{H}' = [\mathbf{H}_1, \mathbf{H}_2, \ldots, \mathbf{H}_K],
\end{equation}
where $\mathbf{H}_k \in \mathbb{C}^{N_r \times N_t}$ represents the downlink channel for the $k$-th subband, with $k = 1, \dots, K$.

Assuming perfect channel estimation at the UE, the normalized eigenvector $\mathbf{w}_k \in \mathbb{C}^{N_t \times 1}$, corresponding to the maximum eigenvalue of the $k$-th subband, is used as the precoding vector \cite{eigennet_liu, eigennet_xiao, mixeddata2024_2}, with $\| \mathbf{w}_k \|^2 = 1$, where $\|\cdot\|$ denotes the $\ell_2$ norm. This vector is obtained through eigenvector decomposition:
\begin{equation}
\mathbf{H}_k^H \mathbf{H}_k \mathbf{w}_k = \lambda_k \mathbf{w}_k,
\end{equation}
where $\lambda_k$ is the largest eigenvalue of $\mathbf{H}_k^H \mathbf{H}_k$, representing the precoding power gain of the MIMO system. For effective downlink precoding, all $K$ eigenvectors must be reported to the BS, leading to the compression and reconstruction of the precoding matrix $\mathbf{W} = [\mathbf{w}_1, \mathbf{w}_2, \ldots, \mathbf{w}_K]$, which contains $K \times N_t$ complex coefficients.

The SGCS metric \cite{3gpp2023csi}, $\rho^2_{k}$, measures the accuracy of CSI feedback and reconstruction for the $k$-th subband:
\begin{equation}
\rho^2_{k} = \frac{||\mathbf{w}_k^H \hat{\mathbf{w}}_k||^2}{\| \mathbf{w}_k \|^2 \| \hat{\mathbf{w}}_k \|^2},
\end{equation}
where $\hat{\mathbf{w}}_k$ is the reconstructed eigenvector for the $k$-th subband. The overall CSI feedback and recovery performance is given by the average SGCS
$
\overline{\rho^2} = \frac{1}{K} \sum_{k=1}^{K} \rho^2_{k}.
$
A value of $\overline{\rho^2}$ closer to 1 indicates higher efficiency in CSI feedback and recovery. Thus, the optimization objective for CSI feedback is defined as:
\begin{equation}
\min _{\mathfrak{J}} -\overline{\rho^2},
\end{equation}
where $\mathfrak{J}$ represents the set of CSI feedback schemes, including codebook-based Type I, Type II, and DL-based methods.

\subsection{DL-based CSI Feedback}

The adaptation of DL-based CSI feedback systems to the diverse and varying characteristics of wireless fading channels requires mobile network operators to collect extensive channel measurements from each environment. These measurements are vital for training DL models to handle the variability in channel conditions effectively. Let $T$ represent the total number of scenarios in a region, where the first $c$ scenarios are used to collect channel samples for training, and the remaining $T - c$ scenarios represent unseen conditions.

We denote the encoding and decoding functions as $f_\text{en}(\cdot)$ and $f_\text{de}(\cdot)$, respectively. To improve the generalization of the CSI feedback network, a common approach is to train the model using a dataset aggregated from multiple scenarios \cite{mixeddata2024, mixeddata2024_2}. After training, the encoder and decoder networks are applied across any scenario as follows:
\begin{eqnarray}
	\mathbf{s} &=& f_{\text{en}}(\mathbf{W};\Phi), \\
	\hat{\mathbf{W}} &=& f_{\text{de}}(\mathbf{s};\Psi),
\end{eqnarray}
where $\Phi$ and $\Psi$ are the parameters of the encoder and decoder networks, respectively.

To further enhance generalization in diverse environments, this paper proposes a preprocessing-assisted optimization framework for CSI feedback. Leveraging knowledge from previously known scenarios, we introduce a standardized input format within the DL feedback framework. All precoding matrices entering the CSI feedback network are first aligned with this standardized format. The standardization process and its inverse are formalized as:
\begin{eqnarray}
	\mathbf{W}_\text{std}, \mathbf{b} &=& \mathcal{S}(\mathbf{W}), \\
	\mathbf{s} &=& f_{\text{en}}(\mathbf{W}_\text{std};\Phi), \\
	\hat{\mathbf{W}}_\text{std} &=& f_{\text{de}}(\mathbf{s};\Psi), \\
	\hat{\mathbf{W}} &=& \mathcal{S}^{-1}(\hat{\mathbf{W}}_\text{std}, \mathbf{b}),
\end{eqnarray}
where $\mathbf{W}_\text{std}$ is the precoding matrix reformatted into the standardized input, and $\mathbf{b}$ encodes the control information detailing the standardization process.  This control information is crucial for restoring the original data format at the BS and is much smaller in size compared to $\mathbf{s}$.

\begin{figure*}[thpb]
      \centering
      \includegraphics[scale=0.26]{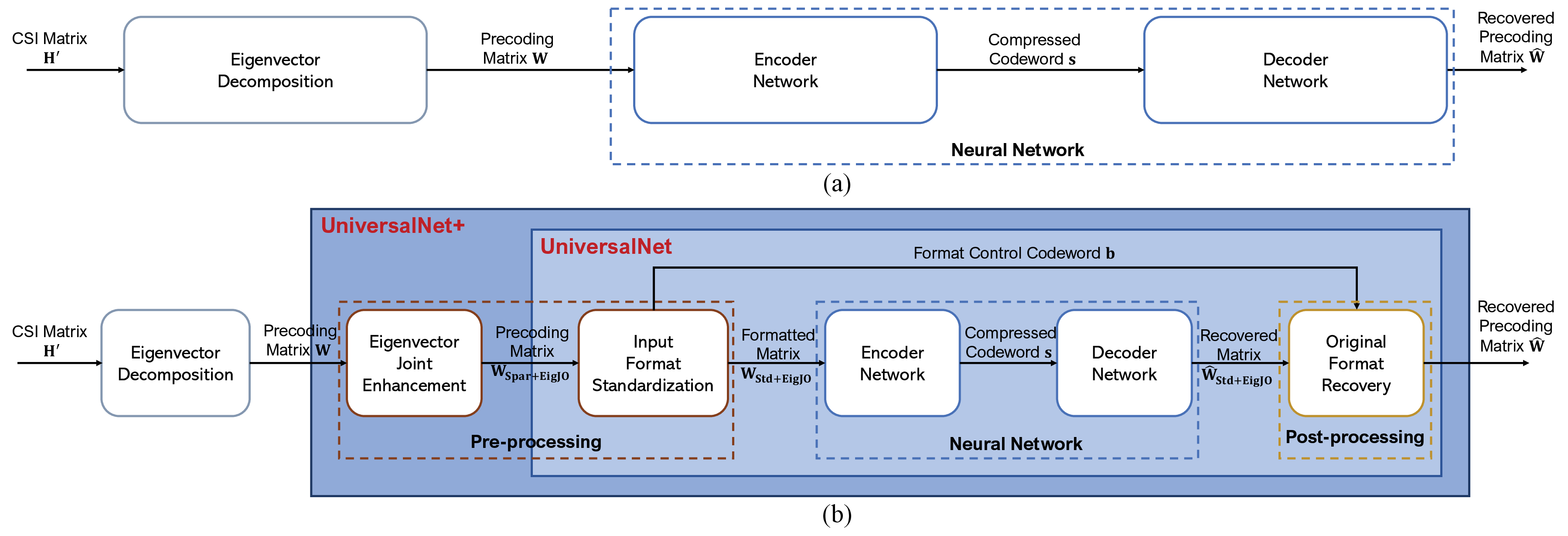}\vspace*{-1mm}
      \caption{Architecture comparison between the conventional DL-based solution (a) and proposed universal solution (b).}
      \label{figure_architecture}
      \vspace*{-5mm}
  \end{figure*}

  \section{Proposed Preprocessing-Based Generalization Enhancement}

Our goal is to develop a universal CSI feedback framework that maintains high generalizability across diverse environments without the need for frequent updates to the backbone model. The conventional DL-based solutions always face the challenge of performance degradation exacerbated by the variability in delay and angle of arrival/departure characteristics across different scenarios, which reduces model effectiveness in new settings, as highlighted by industry studies \cite{3gpp2023csi}. In this section, we propose a preprocessing phase for CSI inputs that standardizes and optimizes the eigenvector, narrowing the environmental variability gap and improving model adaptability.

\subsection{Format Standardization}

Inspired by digital identification practices, where ID photos undergo predefined steps to ensure global applicability and uniqueness, we introduce a framework that applies a standardized input format along with a rapid sparsity-aligning operation. This process reduces environmental variability and enhances compatibility with existing CSI feedback models.

To implement format standardization, the precoding matrix is transformed into a domain with improved sparsity using the 2D Discrete Fourier Transform (DFT). The precoding matrix $\mathbf{W}$ is converted to $\mathbf{W}_\text{Spar}$ as follows:
\begin{equation}
	\mathbf{W}_\text{Spar} = \mathcal{F}(\mathbf{W}) = \mathbf{F}_\text{d}^H \mathbf{W} \mathbf{F}_\text{h},
\end{equation}
where $\mathbf{F}_\text{d}$ and $\mathbf{F}_\text{h}$ are $K \times K$ and $N_t \times N_t$ unitary DFT matrices, respectively. $\mathcal{F}(\cdot)$ denotes the sparse transform function, with 2D DFT chosen due to the inherent sparsity of CSI matrices in the angular-delay domain \cite{cui2022}.

Next, we decouple the phase and magnitude of the precoding matrices during the standardization process:
\begin{equation} 
\mathbf{W}_\text{Spar} = |\mathbf{W}_\text{Spar}| \odot e^{j \angle \mathbf{W}_\text{Spar}}, 
\end{equation}
where $\odot$ denotes the Hadamard product, and $|\cdot|$ and $\angle \cdot$ represent the magnitude and phase of a complex matrix, respectively. The element at position $(i, j)$ in $\mathbf{W}_\text{Spar}$ is expressed as $\mathbf{W}_{i,j} = |\mathbf{W}_{i,j}| e^{j \angle \mathbf{W}_{i,j}}$.

For effective format standardization, the precoding matrices could be aligned to conform to a specified pattern based on minimum path delay, delay spread, minimum angle, and angular spread. This transformation ensures major components align with predefined propagation paths, significantly reducing input data variance for the neural network and enhancing the reuse of pretrained models across varying environments. The trade-off involves additional control information for channel matrix reconstruction at the receiver and the computational complexity of the transformation process.

To minimize the transmission payload and simplify the transformation, we focus on the primary component—either the Line-of-Sight (LoS) path in LoS scenarios or the strongest path in Non-Line-of-Sight (NLoS) scenarios. Given that the main lobe in massive MIMO beamforming predominantly captures the highest power concentration \cite{iddesign_wang}, prioritizing the strongest component is efficient.

A benchmark matrix, $\mathbf{W}_\text{Ben}$, is constructed by selecting a representative LoS channel sample and aligning the major component centrally in the row direction in the angular domain and in the $\lceil \frac{K}{5} \rceil_\text{th}$ column in the delay domain. This accounts for the fact that the strongest LoS component typically represents the shortest path, while in NLoS scenarios, the strongest component is often found within the shorter delay spread range. The magnitude component visualization is shown on the right side of Fig. \ref{figure_visual}.

To align the sparsity characteristics of the precoding matrix $\mathbf{W}$ with the benchmark,  we employ circular shifts defined by the function $f_\text{sh}(\cdot, m, n)$, where $m$ and $n$ represent the shift steps in rows and columns, respectively. The shifted precoding matrix $\mathbf{W}_\text{shift}$ is defined as:
\begin{equation}
    \mathbf{W}_\text{shift} = f_\text{sh}(\mathbf{W}_\text{Spar}, m, n) = \mathbf{A}^m \mathbf{W}_\text{Spar} (\mathbf{A}^T)^n,
\end{equation}
where $\mathbf{A} \in \mathbb{R}^{N_t \times N_t}$ is the cyclic permutation matrix:
\begin{equation}
    \mathbf{A} = \begin{pmatrix}
    0 & \cdots & 0 & 1 \\
    1 & \cdots & 0 & 0 \\
    \vdots & \ddots & \vdots & \vdots \\
    0 & \cdots & 1 & 0 \\
    \end{pmatrix},
\end{equation}
and $\mathbf{A}^T$ is its transpose. The matrices $\mathbf{A}$ and $\mathbf{A}^T$ facilitate cyclic row and column permutations of $\mathbf{W}_\text{Spar}$.

Subsequently, our goal is to determine the optimal shift steps $m^*$ and $n^*$ that best align $\mathbf{W}_\text{Spar}$ with the benchmark  $\mathbf{W}_\text{Ben}$.

To simplify computation, we solve for $m^*$ and $n^*$ independently. We define the row sum vector:
\begin{equation}
    \mathbf{r} = \left[ \sum_{j} |\mathbf{W}_{1,j}|, \sum_{j} |\mathbf{W}_{2,j}|, \ldots, \sum_{j} |\mathbf{W}_{K,j}| \right]^T,
\end{equation}
and similarly define $\mathbf{r}_\text{Ben}$ for the benchmark matrix $|\mathbf{W}_\text{Ben}|$.

We identify the optimal cyclic shift $m^*$ that maximizes the correlation between $\mathbf{r}$ and $\mathbf{r}_\text{Ben}$ using the correlation function:
\begin{equation}
    C(m) = \mathbf{r}_m \cdot \mathbf{r}_\text{Ben},
\end{equation}
where $\mathbf{r}_m$ is $\mathbf{r}$ after a cyclic shift by $m$ steps. The optimal shift $m^*$ is found by:
\begin{equation}
    m^* = \arg\max_m C(m).
\end{equation}

A similar process applies to finding the optimal column shift $n^*$, with the column sum vector defined as:
\begin{equation}
    \mathbf{c} = \left[ \sum_{i} |\mathbf{W}_{i,1}|, \sum_{i} |\mathbf{W}_{i,2}|, \ldots, \sum_{i} |\mathbf{W}_{i,N_t}| \right]^T,
\end{equation}
and seeking the shift that maximizes the correlation with $\mathbf{c}_\text{Ben}$ defined for the benchmark matrix.

The standardized precoding matrix $\mathbf{W}_\text{Std}$ is obtained as:
\begin{equation}
    \mathbf{W}_\text{Std} = \mathbf{A}^{m^*} \mathbf{W}_\text{Spar} (\mathbf{A}^T)^{n^*},
\end{equation}
where $\{m^*, n^*\}$ are the optimal shift steps included in the control information $\mathbf{b}$. To encode this control information, $B_\text{ctrl} = \lceil \log_2{K} \rceil + \lceil \log_2{N_t} \rceil$ bits are necessary, where $\lceil \cdot \rceil$ is the ceiling function.

To reconstruct the precoding matrix from the neural network's output $\hat{\mathbf{W}}_\text{Std}$, we apply the inverse of the format standardization process:
\begin{equation}
    \hat{\mathbf{W}} = \mathcal{F}^{-1}\left( \mathbf{A}^{-m^*} \hat{\mathbf{W}}_\text{Std} (\mathbf{A}^T)^{-n^*} \right),
\end{equation}
where $\mathcal{F}^{-1}(\cdot)$ denotes the inverse sparse transform function, specifically the inverse 2D DFT, enabling recovery of the original precoding matrix format.

\subsection{Eigenvector Joint Optimization}

In this subsection, we jointly optimize the eigenvectors of all subbands in the downlink channel to enhance the sparsity of the precoding matrix, thereby enabling high-efficiency compression.

Given a specific eigenvalue $\lambda_k$ for subband $k$, there exist infinitely many eigenvectors $\mathbf{w}_k \in \mathcal{E}_k$ that satisfy Eq.~(2), where $\mathcal{E}_k$ denotes the corresponding eigenspace. However, randomness in these eigenvectors disrupts the consistency of physical environment information within the precoding matrix, exacerbating sparsity challenges and complicating compression. Consequently, it is essential to enhance sparsity in the transformed domain while preserving precoding performance. We aim to find adjusted eigenvectors $\mathbf{W}_{\text{EigJO}} = [\mathbf{\tilde{w}}_1,\, \mathbf{\tilde{w}}_2,\, \ldots,\, \mathbf{\tilde{w}}_K]$ that provide optimal sparsity for efficient compression:
\begin{equation}
[\mathbf{\tilde{w}}_1,\, \mathbf{\tilde{w}}_2,\, \ldots,\, \mathbf{\tilde{w}}_K] = \underset{\mathbf{w}_k \in \mathcal{E}_k,\, k = 1, 2, \ldots, K}{\operatorname{arg\,min}} \left\| \mathcal{F}([\mathbf{w}_1,\, \mathbf{w}_2,\, \ldots,\, \mathbf{w}_K]) \right\|_1,
\end{equation}
where $\|\cdot\|_1$ denotes the $\ell_1$-norm, and $\mathcal{F}(\cdot)$ represents the transform into the angular-delay domain using DFT.

Due to computational complexity limitations in CSI feedback, we adopt a suboptimal but efficient approach for enhancing the sparsity of the precoding matrices. Given that the channel matrix $\mathbf{H}' = [\mathbf{H}_1,\, \mathbf{H}_2,\, \ldots,\, \mathbf{H}_K]$ exhibits notable sparsity in the angular-delay domain due to spatial and spectral correlation~\cite{Overview}, we propose to calculate eigenvectors that align more closely with the channel vectors in their respective subbands.

Specifically, we aim to find adjusted eigenvectors $\mathbf{\tilde{W}}$ that are as close as possible to reference vectors derived from the channel matrices while remaining within the eigenspaces corresponding to their respective eigenvalues. This can be formulated as:
\begin{equation}
\mathbf{\tilde{w}}_k = \underset{\mathbf{w}_k \in \mathcal{E}_k}{\operatorname{arg\,min}} \left\| \mathbf{w}_k - \mathbf{v}_k \right\|, \quad k = 1, 2, \ldots, K,
\end{equation}
where $\|\cdot\|$ denotes the $\ell_2$-norm, $\mathcal{E}_k$ is the eigenspace corresponding to the eigenvalue $\lambda_k$, and $\mathbf{v}_k$ is a reference vector derived from the channel matrix $\mathbf{H}_k$.

Without loss of generality, we choose $\mathbf{v}_k$ to be the transpose of the first row of the channel matrix $\mathbf{H}_k \in \mathbb{C}^{N_r \times N_t}$ in the $k$-th subband; that is, $\mathbf{v}_k = (\mathbf{H}_k)_{1,:}^T \in \mathbb{C}^{N_t \times 1}$, representing the channel vector between the first receive antenna and all transmit antennas. Note that the channel vector between any other receive antenna and all transmit antennas can also be chosen as $\mathbf{v}_k$.

The adjusted eigenvector $\mathbf{\tilde{w}}_k$ is then obtained by projecting $\mathbf{v}_k$ onto the eigenspace $\mathcal{E}_k$:
\begin{equation}
\mathbf{\tilde{w}}_k = \mathbf{P}_k\mathbf{v}_k,
\end{equation}
where $\mathbf{P}_k$ is the projection operator onto $\mathcal{E}_k$:
\begin{equation}
\mathbf{P}_k = \mathbf{E}_k\mathbf{E}_k^H,
\end{equation}
and $\mathbf{E}_k$ is the matrix whose columns form an orthonormal basis for $\mathcal{E}_k$. Finally, by normalizing $\mathbf{\tilde{w}}_k$, we ensure that the adjusted eigenvector has unit norm:
\begin{equation}
\mathbf{\tilde{w}}_k = \frac{\mathbf{\tilde{w}}_k}{\left\| \mathbf{\tilde{w}}_k \right\|}.
\end{equation}

This projection ensures that $\mathbf{\tilde{w}}_k$ retains the eigenvector property with respect to $\mathbf{H}_k^\mathrm{H} \mathbf{H}_k$ and aligns closely with the channel characteristics, enhancing sparsity in the angular-delay domain.

The benchmark matrix $\mathbf{W}_\text{Ben+EigJO}$ and the general sparse precoding matrix $\mathbf{W}_\text{Spar+EigJO}$ after eigenvector joint optimization are obtained through Eqs.~(23)--(25). Note that $\mathbf{W}_\text{Ben+EigJO}$ needs to be computed only once during initialization.

With eigenvector joint optimization, the final standardized input matrix is:
\begin{equation}
\mathbf{W}_\text{Std+EigJO} = \mathbf{A}^{m^*} \mathbf{W}_\text{Spar+EigJO} \left( \mathbf{A}^\mathrm{T} \right)^{n^*}.
\end{equation}
\vspace*{-6.5mm}

\begin{algorithm}[ht]
\caption{Preprocessing-Based Generalization Enhancement Algorithm}
\begin{algorithmic}[1]
\Statex \textbf{Initialization:} Number of transmit antennas $N_t$, number of receive antennas $N_r$, number of subbands $K$, benchmark matrix $\mathbf{W}_{\text{Ben+EigJO}}$;
\Statex \textbf{Input:} New channel sample $\mathbf{H}' = [\mathbf{H}_1,\, \mathbf{H}_2,\, \ldots,\, \mathbf{H}_K]$;
\For{$k = 1$ to $K$}
    \State \parbox[t]{\dimexpr\linewidth-\algorithmicindent}{Perform eigenvalue decomposition of $\mathbf{H}_k^{H} \mathbf{H}_k$ to obtain eigenvalues $\lambda_k$ and eigenspaces $\mathcal{E}_k$;\strut}
    \State \parbox[t]{\dimexpr\linewidth-\algorithmicindent}{Compute the adjusted eigenvector $\mathbf{\tilde{w}}_k$ by projecting $\mathbf{v}_k$ onto $\mathcal{E}_k$ using Eqs.~(23)--(25);\strut}
\EndFor
\State Assemble the optimized precoding matrix $\mathbf{W}_{\text{EigJO}} = [\mathbf{\tilde{w}}_1,\, \mathbf{\tilde{w}}_2,\, \ldots,\, \mathbf{\tilde{w}}_K]$ and calculate $\mathbf{W}_{\text{Spar+EigJO}}$ using the sparse transform $\mathcal{F}(\cdot)$;
\State Determine the optimal shift lengths $\{ m^*,\, n^* \}$ to align with $\mathbf{W}_{\text{Ben+EigJO}}$ as per steps (15)--(18);
\State Compute the final standardized input matrix $\mathbf{W}_{\text{Std+EigJO}}$ using Eq.~(26);
\Statex \textbf{Output:} Control information $\mathbf{b} = \{ m^*, n^* \}$ for feedback, standardized precoding matrix $\mathbf{W}_{\text{Std+EigJO}}$ for compression and feedback.
\end{algorithmic}
\end{algorithm}

\subsection{Visualization of the Proposed Preprocessing Method}

\begin{figure}[thpb]
    \centering
    \includegraphics[scale=0.30]{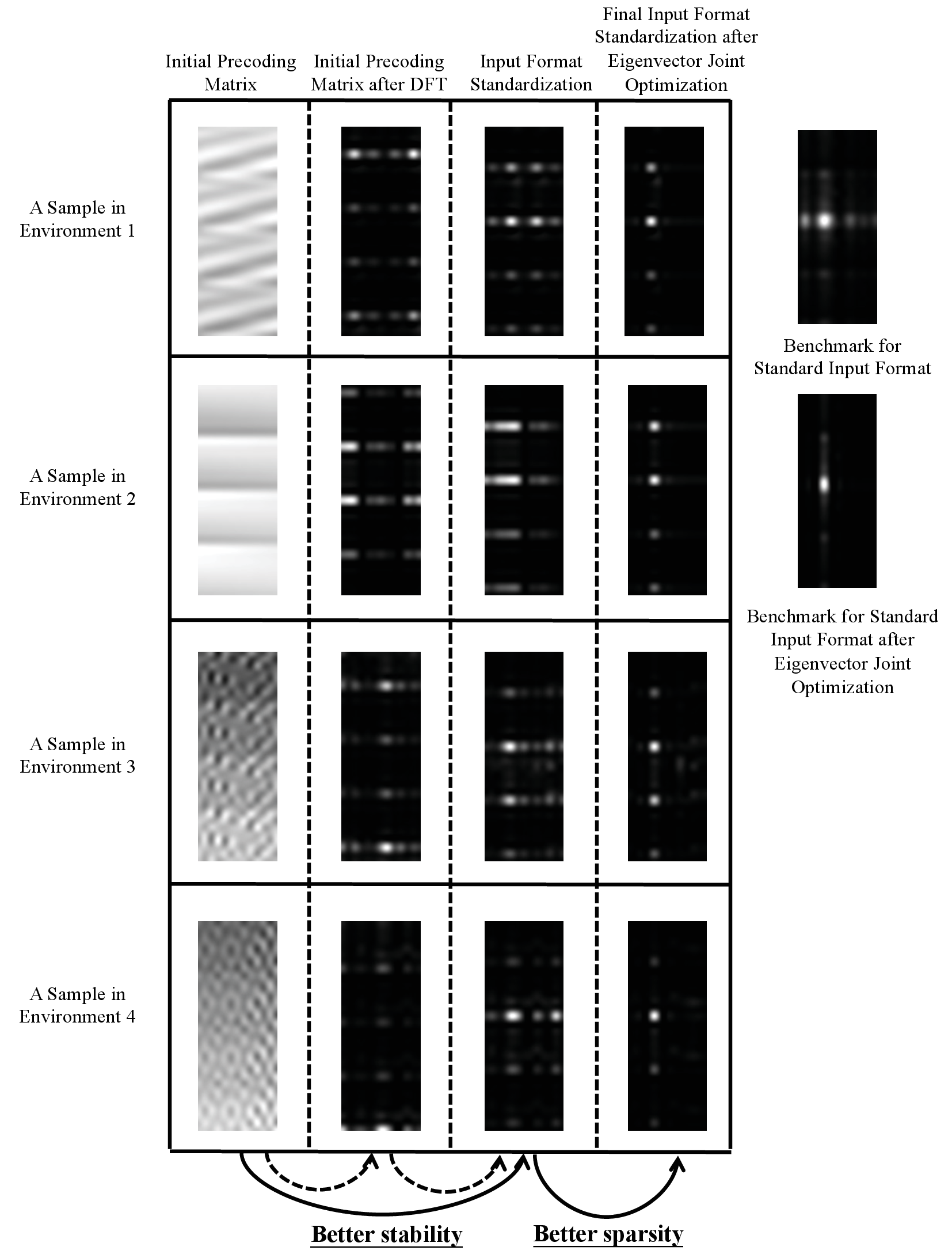}\vspace*{-1mm}
    \caption{Visualization of precoding matrix magnitudes for different environments using proposed preprocessing methods.}
    \label{figure_visual}
\end{figure}

Fig. \ref{figure_visual} illustrates the impact of the proposed preprocessing stages on precoding matrices derived from samples in four different environments. In these visualizations, the brightness intensity corresponds to the magnitude of the matrix elements, serving as a visual indicator of sparsity and element strength.

Initially, the precoding matrices exhibit significant variations in element magnitudes, reflecting the unique channel characteristics of each environment. After applying the DFT, although the environmental differences persist, the matrices become sparser, highlighting the DFT's role in restructuring the matrix to reveal inherent sparsity patterns.

The subsequent application of input format standardization further homogenizes the matrices across different environments, as seen in the third column of Fig.~\ref{figure_visual}. This step introduces a regular sparsity pattern, enhancing uniformity and facilitating more efficient compression and consistent precoding performance regardless of the underlying channel conditions.

Finally, integrating eigenvector joint optimization alongside input format standardization markedly amplifies the sparsity of the matrices, as shown in the last column. This enhancement is visually evident through intensified bright spots within a darker background, indicating reduced randomness and increased concentration of element magnitudes. This pattern demonstrates the optimization's effectiveness in mitigating the adverse effects of eigenvector randomness on sparsity in the transformed domain, thereby improving compression efficiency.

Overall, the sequential visualizations underscore the effectiveness of the proposed eigenvector optimization and standardization methods in promoting model generalization. By establishing a standardized input format and enhancing sparsity, these methods significantly contribute to the robust adaptation of compression models across various wireless communication environments.
\subsection{Encoder and Decoder Networks}

The encoder and decoder networks can be chosen flexibly, allowing for any DL-based CSI feedback models, such as those detailed in \cite{cui2022, mourya2023, eigennet_liu, mixeddata2024_2}. According to 3GPP TR~38.843 \cite{3gpp2023csi}, industry leaders such as Nokia, Huawei, Qualcomm, Apple, and OPPO predominantly utilize Transformer models for CSI feedback. Specifically, TransNet \cite{cui2022} employs a two-layer Transformer architecture that integrates an attention mechanism to learn interconnections within CSI data during feedback. Given that TransNet's source code is available on GitHub, we use it as our primary example to demonstrate the effectiveness of the proposed methods. Additionally, we adopt the quantization approach described in \cite{lu2020bit} for quantifying the dimension-compressed floating-point vector.

The total feedback bit count, $B_\text{total}$, for an encoder output dimension of $L$ and $B$-bit quantization per codeword element, supplemented by control bits $B_\text{ctrl}$, is calculated as
$
B_\text{total} = L \times B + B_\text{ctrl},
$
where $B_\text{ctrl} \ll B_\text{total}$ in general.

\section{Performance Evaluation}

\subsection{Experiment Setup}

The performance is evaluated using a ray-tracing-based wireless channel simulation platform \cite{WirelessAI2022}, which generates CSI data from 100 real-world maps of major cities worldwide. Each map provides detailed channel information between a BS and 10,000 UE locations, reflecting realistic building layouts. The fundamental simulation parameters are summarized in Table~I.
\vspace*{-3mm}

\begin{table}[h!]
\centering
\caption{Basic Simulation Parameters}
\begin{tabular}{|l|c|}
\hline
\textbf{Parameter} & \textbf{Value} \\
\hline
System bandwidth & 10 MHz \\
\hline
Carrier frequency & 2.6 GHz \\
\hline
Subcarrier spacing & 15 kHz \\
\hline
Number of subcarriers \(N_{sc}\) & 624 \\
\hline
Subband granularity \(N_{gran}\) & 48 \\
\hline
Number of subbands \(N_{sb}\) & 13 \\
\hline
Horizontal Tx antenna ports per polarization \(N_h\) & 8 \\
\hline
Vertical Tx antenna ports per polarization \(N_v\) & 4 \\
\hline
Polarization type & Single \\
\hline
Total Tx antenna ports \(N_t\) & 32 \\
\hline
Rx antennas \(N_r\) & 4 \\
\hline
\end{tabular}
\end{table}

We use a training dataset of 160,000 samples and a testing set of 60,000 samples, including 2,000 samples from 30 unseen environments. To examine the impact of mixed datasets, the training samples are sourced from the first 1 to 70 maps. If the initial dataset is insufficient, it is augmented through repetition to match the desired size; if too large, it is uniformly sampled.  The models are trained with a batch size of 64. Specifically, TransNet undergoes 1,600 training epochs, followed by 600 epochs of fine-tuning with the quantization module. The quantization bits per element are set to $B =$ 6, and the control bits are $B_\text{ctrl} =$ 9.

\subsection{Performance Over Various Feedback Bits and Training Environments}

\begin{figure} \hspace*{-3mm}
\subfigure[1 environment] {\label{figpe2:a} 
\includegraphics[width=0.51\columnwidth]{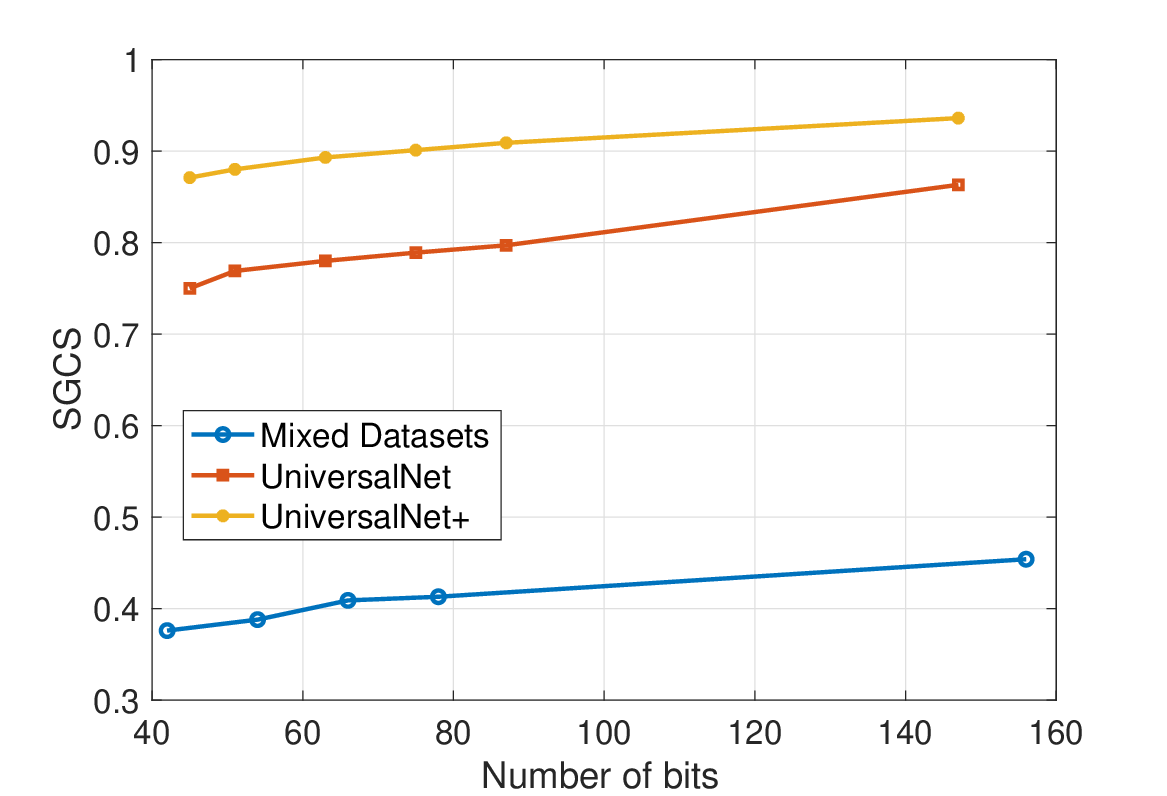}
} \hspace*{-8mm}
\subfigure[4 environments] { \label{figpe2:b} 
\includegraphics[width=0.51\columnwidth]{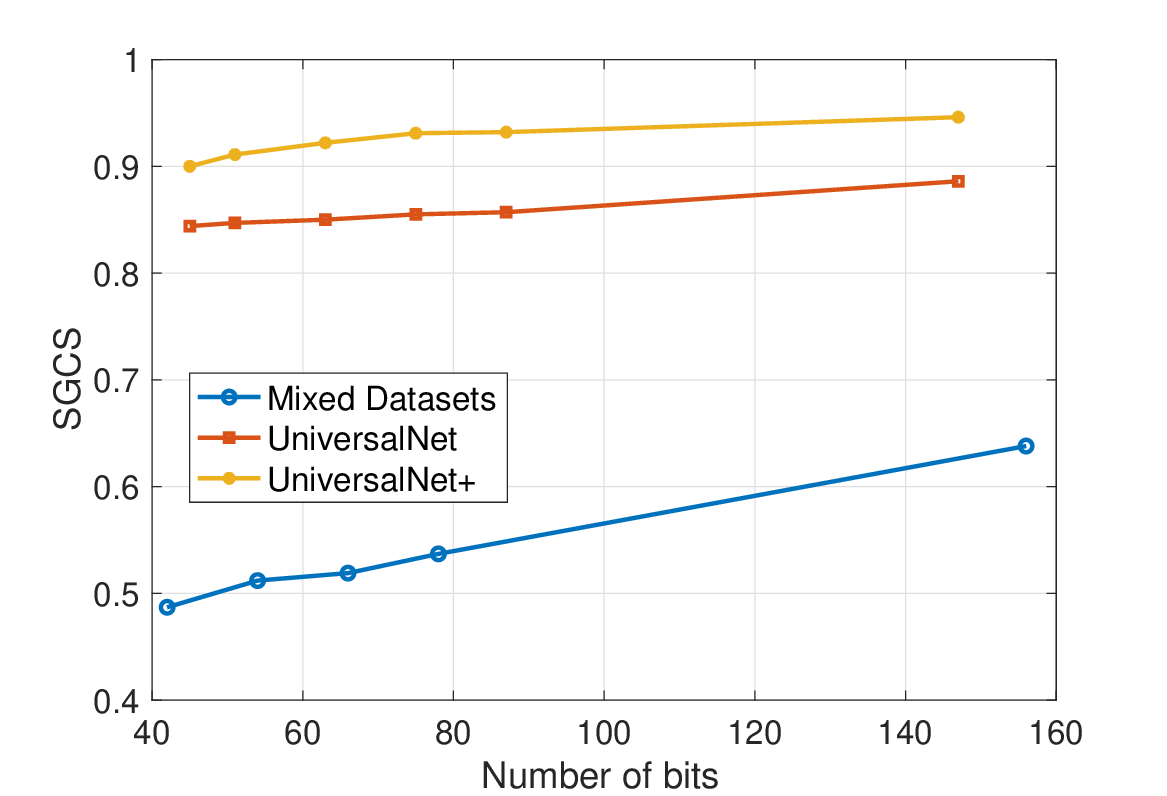} 
} \hspace*{-3mm}
\subfigure[16 environments] {\label{figpe2:c} 
\includegraphics[width=0.51\columnwidth]{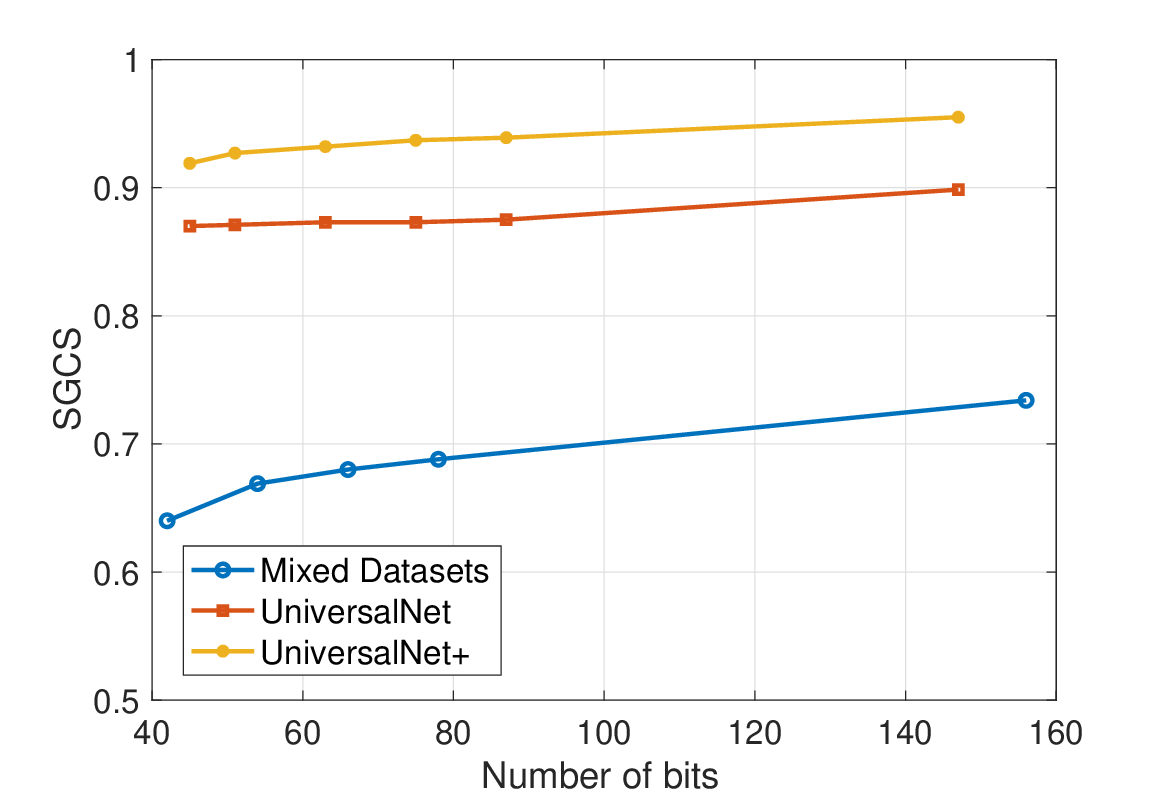}
} \hspace*{-8mm}
\subfigure[70 environments] { \label{figpe2:d} 
\includegraphics[width=0.51\columnwidth]{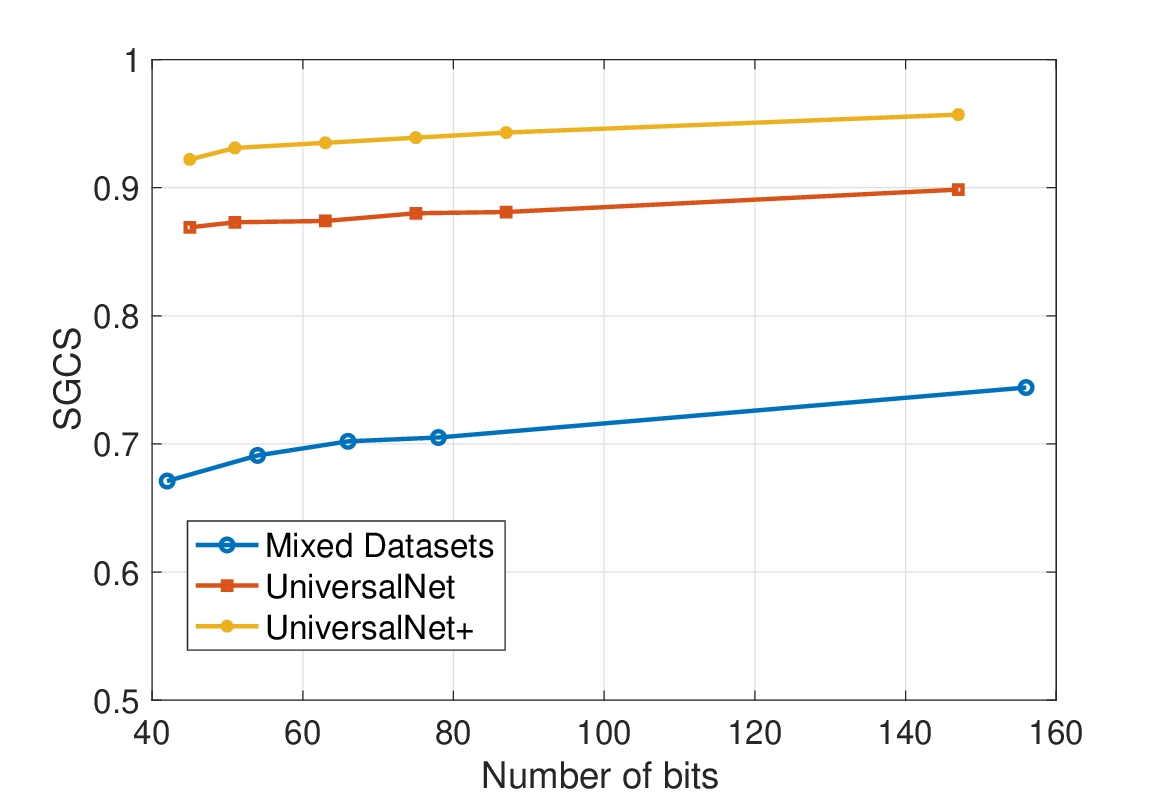} 
} 
\vspace*{-3mm}
\caption{CSI recovery comparison in the unseen environments at different feedback bits when the training set is collected from the given number of environments.} 
\vspace*{-5mm}
\label{figurepe2} 
\end{figure}

Fig.~\ref{figurepe2} illustrates the impact of our generalization enhancement techniques on CSI recovery in unseen environments, considering varying numbers of feedback bits and training environments. The method denoted as ``UniversalNet+'' represents the performance of the transformer-based CSI feedback network \cite{cui2022}, augmented with our proposed input format standardization and joint eigenvector optimization. In contrast, ``UniversalNet'' reflects outcomes without joint eigenvector enhancement, and ``Mixed Datasets'' refers to the baseline transformer-based CSI feedback network that employs mixed datasets \cite{mixeddata2024, mixeddata2024_2} without our preprocessing interventions.

As shown in Fig.~\ref{figurepe2}, the ``UniversalNet+'' method demonstrates superior generalization capabilities across all scenarios, confirming the effectiveness of input data format standardization and joint eigenvector optimization in enhancing CSI feedback in unseen environments. Notably, even without joint eigenvector optimization, ``UniversalNet'' shows significant improvement over the ``Mixed Datasets'' method, highlighting the benefits of our standardization design inspired by digital identification processes.

\begin{figure}[!t]
\centering
\includegraphics[width=0.98\columnwidth]{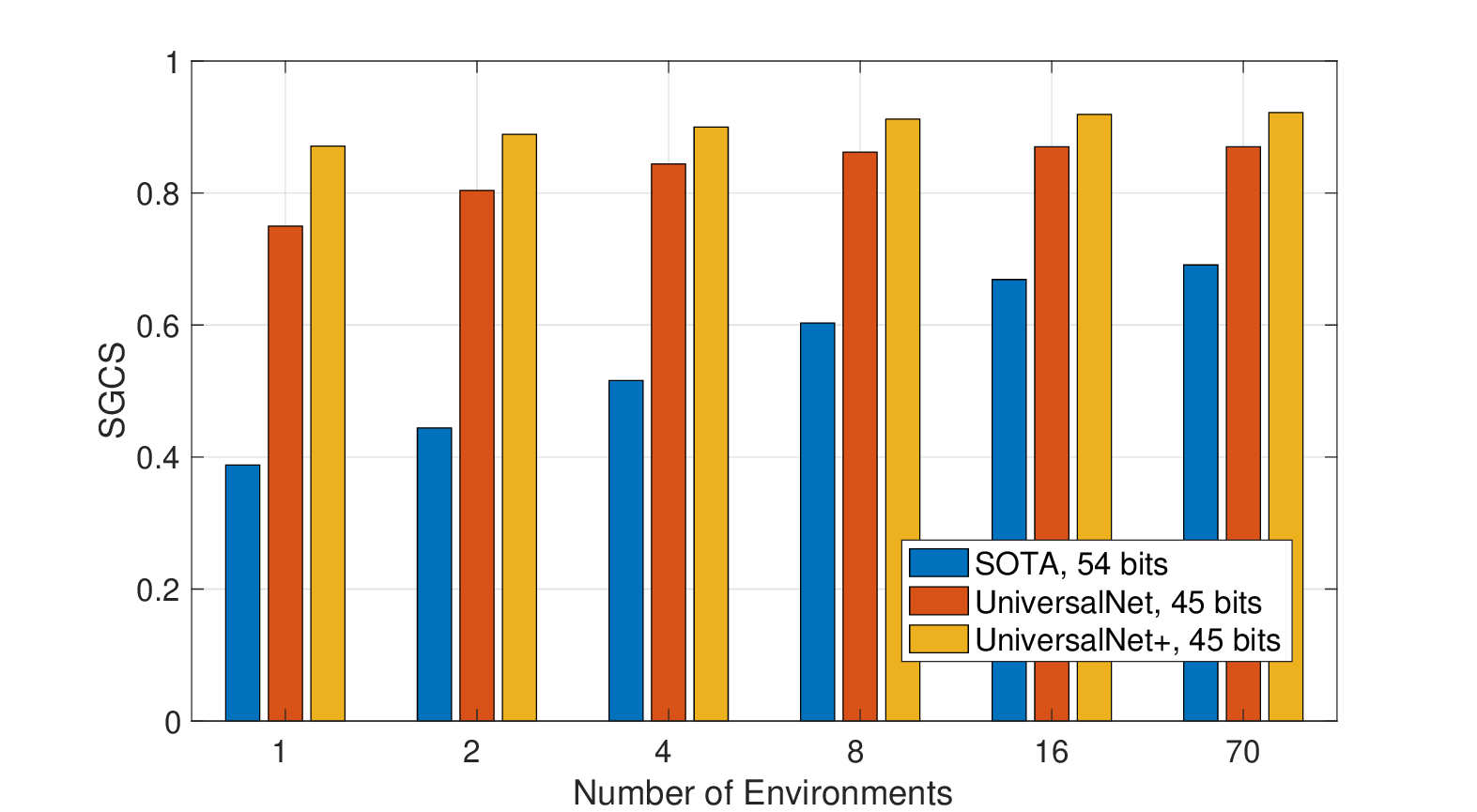}
\vspace*{-2mm}
\caption{Detailed CSI recovery comparison across a varying number of training environments, with feedback bits set around 45.}
\vspace*{-6mm}
\label{figure_envs}
\end{figure}

Fig.~\ref{figure_envs} further illustrates the performance at approximately 45 feedback bits, reinforcing the insights from Fig.~\ref{figurepe2}. The ``UniversalNet+'' model consistently outperforms both the baseline and the ``UniversalNet'' configurations in unfamiliar environments. Remarkably, a model trained with data from a single map can attain an SGCS of nearly 0.9 with 45 feedback bits across 30 unseen maps. The sustained superior performance across a growing number of environments validates the proposed eigenvector optimization technique's ability to provide a generalized and efficient CSI feedback solution suitable for practical deployment in varied and dynamically changing wireless communication scenarios.

\section{Conclusions}
In this paper, we tackle the pressing issue of generalization in DL-based CSI feedback mechanisms within wireless communication systems through an innovative approach inspired by digital identification processes. By introducing a preprocessing phase that standardizes CSI inputs and optimizes eigenvector preprocessing, our methodology diverges from conventional neural network optimizations, offering a novel avenue for enhancing the adaptability of CSI feedback solutions across varied RF environments. The establishment of a standardized input format specifically addresses the challenges of environmental variability, leading to significant improvements in codeword compression efficiency and, ultimately, the robustness of wireless communication systems.


\ifCLASSOPTIONcaptionsoff
  \newpage
\fi

\bibliographystyle{IEEEtran}
\bibliography{ref}

\end{document}